\documentclass[twocolumn,showpacs,preprintnumbers,reprint,amsmath,amssymb,aps,superscriptaddress,float]{revtex4-1}
\usepackage{graphicx}
\usepackage{dcolumn}
\usepackage{bm}
\usepackage{txfonts}
\usepackage{color}
\usepackage[colorlinks,citecolor=blue]{hyperref}

\begin{document}
\title{High-energy-density \emph{$e^-e^+$} pair plasma production and its dynamics in the \\relativistic transparency regime}
\date{\today}

\author{W. Y. Liu}
\affiliation{School of Nuclear Science and Technology, University of South China, Hengyang 421001, China}
\affiliation{Key Laboratory for Laser Plasmas (Ministry of Education),School of Physics and Astronomy,\\ Shanghai Jiao Tong University, Shanghai 200240, China}
\affiliation{IFSA Collaborative Innovation Center, Shanghai Jiao Tong University, Shanghai 200240, China}
\author{W. Luo}
\email{wenluo-ok@163.com}
\affiliation{School of Nuclear Science and Technology, University of South China, Hengyang 421001, China}
\affiliation{SUPA, Department of Physics, University of Strathclyde, Glasgow G40NG, United Kingdom}
\author{T. Yuan}
\affiliation{Key Laboratory for Laser Plasmas (Ministry of Education),School of Physics and Astronomy,\\ Shanghai Jiao Tong University, Shanghai 200240, China}
\affiliation{IFSA Collaborative Innovation Center, Shanghai Jiao Tong University, Shanghai 200240, China}
\author{J. Y. Yu}
\affiliation{Key Laboratory for Laser Plasmas (Ministry of Education),School of Physics and Astronomy,\\ Shanghai Jiao Tong University, Shanghai 200240, China}
\affiliation{IFSA Collaborative Innovation Center, Shanghai Jiao Tong University, Shanghai 200240, China}
\author{M. Chen}
\email{minchen@sjtu.edu.cn}
\affiliation{Key Laboratory for Laser Plasmas (Ministry of Education),School of Physics and Astronomy,\\ Shanghai Jiao Tong University, Shanghai 200240, China}
\affiliation{IFSA Collaborative Innovation Center, Shanghai Jiao Tong University, Shanghai 200240, China}
\author{Z. M. Sheng}
\affiliation{Key Laboratory for Laser Plasmas (Ministry of Education),School of Physics and Astronomy,\\ Shanghai Jiao Tong University, Shanghai 200240, China}
\affiliation{IFSA Collaborative Innovation Center, Shanghai Jiao Tong University, Shanghai 200240, China}
\affiliation{SUPA, Department of Physics, University of Strathclyde, Glasgow G40NG, United Kingdom}

\begin{abstract}
High-energy-density electron-positron (\emph{$e^-e^+$}) pair plasma production and its dynamics in a thin foil illuminated by two counter-propagating laser pulses are investigated through multi-dimensional particle-in-cell simulations. We compare the production of \emph{$e^-e^+$} pairs and $\gamma$-photons via quantum electrodynamics processes in the relativistic transparent and opaque regimes, and find that the target transparency can significantly enhance the \emph{$e^-e^+$} pair production due to the formation of stable standing wave (SW). An optimum foil density of 200~--~280$~n_c$ ($n_c$ is the laser critical density) is found for enhancing \emph{$e^-e^+$} pair production when laser intensity reaches a few $10^{23}~\rm{W/cm^2}$. At such foil density, laser energy conversion to \emph{$e^-e^+$} pairs is approximately four times higher than at foil density of 710$~n_c$, whereas laser energy conversion to $\gamma$-photons keeps almost the same. Consequently high dense \emph{$e^-e^+$} plasma with a maximum intensity above $10^{20}~\rm{W/cm^2}$ is produced. Modulation dynamics of created pair plasmas is further observed when target foil becomes transparent. It is shown that stable SWs formed directly by two counter-propagating lasers, not only trap the created \emph{$e^-e^+$} pairs to their nodes, but also modulate periodically average energy and phase-space and angular distributions of trapped particles. However, similar trapping and modulation effects become obscure in the opaque regime due to the absence of stable SW field.
\end{abstract}

\maketitle

\section{Introduction}
\vspace{-0.5 em}
Relativistic electron-positron (\emph{$e^-e^+$}) pair plasma represents a unique state of matter, which is believed to exist in many extreme astrophysical environments, such as Active Galactic Nuclei (AGN), Pulsar Wind Nebulae (PWN), Gamma Ray Bursts (GRBs), and Black Holes (BHs) \cite{E. Weibel1959,B. Paczynski1986,M. J. Rees1992,E. Waxman1995,T. Piran1999,T. Piran2005,P. Mészáros2006,Hui Chen pop2015}. In particular, high dense \emph{$e^-e^+$} pair plasmas are generally accepted to be invoked to explain energetic phenomena associated with quasars, GRBs and BHs \cite{J. Wardle1998,I. F. Mirabel1999,P. Meszaros2002,G. Weidenspointner2008,G. Sarri2015}. On earth, relativistic positron sources are of paramount importance in experimental physics due to their potential applications to a wide range of physical subjects, including nuclear physics, particle physics, and laboratory astrophysics \cite{wluo2013,nd2016}. Generation of such pair plasma is attracting an increasing amount of worldwide attention \cite{H. Chen2011,Hui Chen prl2015,Jian-Xun Liu2015,Jin-Jin Liu2016,Tongjun Xu2016,yuantao2017}, aiming to replicate the physics of small-scale astrophysical environments in the laboratory and to elaborate potentially board applications.

With everlasting and progressive development of laser technology, laser beams with intensities of $10^{22}~\rm{W/cm^2}$ are already realized \cite{V. Yanovsky2008}. The underway project Extreme Light Infrastructure (ELI) \cite{eli} and proposed initiatives like XCELS \cite{xcels} and iCAN \cite{G. Mourou2013} are expected to achieve intensities above $10^{23}~\rm W/cm^2$ within the next few years. At such intensities, the normalized amplitude of laser vector potential $a_0=eE/(m_ec\omega_l)\gg1$, where $\omega_l=2\pi c/\lambda_l$ is the angular frequency of the laser, and laser-solid interaction enters the ultrarelativistic regime \cite{L.Willingale2009}, in which quantum electrodynamics (QED) processes may arise \cite{G. A. Mourou2006,E. N. Nerush2007,A. R. Bell2008,A. M. Fedotov2010}. One of the most representative QED processes is nonlinear Compton Scattering, $e^-+n\gamma_l\rightarrow e^-+\gamma_h$ \cite{A. DiPiazza2010,F. Mackenroth2011,C. P. Ridgers2012}. In this nonlinear process, foil electron accelerated (in the laser focus) spontaneously absorbs multiple laser photons $\gamma_l$ and then radiate a high-energy photon $\gamma_h$. The radiated photon in strong electromagnetic (EM) fields can further decay into an \emph{$e^-e^+$} pair via multi-photon Breit-Wheeler (BW) process, $\gamma_h+m\gamma_l\rightarrow e^-+e^+$ \cite{G. Breit1934,T. Erber1966,J. G. Kirk2009}. As QED effects become significant, prolific charged particles produced in the system can focus large current into the classical electron-ion plasma, which strongly modify the plasma dynamics. The EM fields, which are determined by the plasma dynamics, in turn affect the rates of the QED processes. As a result, the physics associated with the formation and dynamics of dense \emph{$e^-e^+$} pair plasmas produced by QED-strong laser interacting with solid foil is of paramount importance that may prove essential for developing an astrophysical relevant research platform.

\begin{figure*}
  \centering
  \includegraphics[scale=0.35]{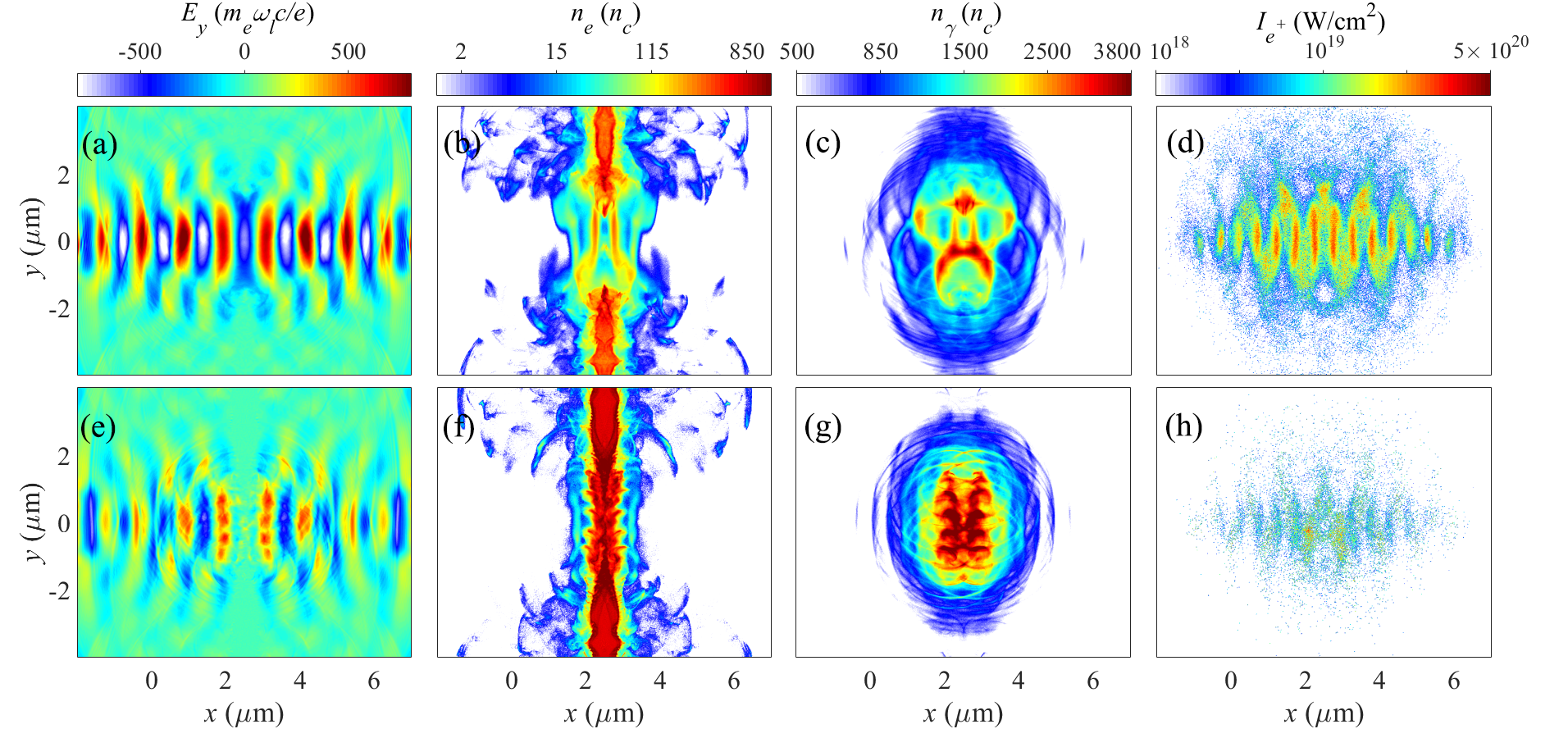}
  \caption{(Color online) The $y$-component of the normalized electric fields $E_y$ [(a) and (e)], the spatial density maps in units of $n_c$ of the foil electrons [(b) and (f)], and emitted $\gamma$-ray photons [(c) and (g)], and the positron intensity averaged one laser period [(d) and (h)] for a laser intensity of $I=4\times10^{23}~\rm{W/cm^2}$ ($a_0=540$). The data was recorded at $t=10~T_0$ and the initial target densities are 280$~n_c$  (upper pads) and 710$~n_c$ (bottom pads), respectively.}\label{fig1}
\end{figure*}

In this paper, we present enhanced \emph{$e^-e^+$} pair plasma production and its dynamics in thin foils undergoing relativistic transparency. The case of a thin foil irradiated by two counter-propagating lasers (two-side irradiation scheme) is particularly interesting, because in contrast to single laser solid interaction, two-side irradiation scheme can significantly enhance the QED effects \cite{Wen Luo2015,H. X. Chang2015}. Here we revisit this irradiation scheme in the relativistic transparency regime and emphasize on how this affects the pair production and the following particle dynamics. When target foil undergoes transparency, stable standing-wave (SW) field can be produced directly by the overlap of two counter-propagating laser pulses. Such SW fields significantly enhance the pair production by the BW process and then produce high dense \emph{$e^-e^+$} pair plasmas. Modulation dynamics of the positron energy and phase-space and angular distributions is further observed when transparency occurs.

The reminder of this paper consists of five sections. Section II describes the simulation setup. Section III demonstrates dense \emph{$e^-e^+$} pair plasma and $\gamma$-ray burst production in both the relativistic transparency and opaque regimes. In Section IV, the modulation of stable SW fields on produced pair plasmas in the relativistic transparency regime is discussed. Section V presents a brief discussion and summary on this work.
\vspace{-4.5 em}
\section{simulation setup}
\vspace{-1 em}
An open-source particle-in-cell (PIC) code EPOCH \cite{C. P. Ridgers2014} was used to perform multi-dimensional simulations. This code has been incorporated with a radiation-reaction module and a collective QED module, allowing self-consistent modeling of laser-plasmas interactions in the near-QED regime. Two linearly polarized (LP) laser pulses with an equal wavelength of $\lambda_l=1~\mu m$ and duration of $\tau=9~T_0$ focus on, respectively, the left and right boundaries of the simulation box at time $t=0$. Here $T_0=\lambda_l/c$ is the period of laser pulse. The lasers have transversely super-Gaussian spatial profiles with electric field as $\varpropto exp(-y^5/1~\mu m^2)$ and focus to spots with radius of $r = 1~\mu m$. The size of simulation box is $9\lambda_l\times8\lambda_l$. A thin foil is fully ionized carbons and hydrogen, which locates between $4~\mu m$ to $5~\mu m$ from the left boundary. Ratio of density of carbons and hydrogen is 1:1. The foil is discretized on a spatial grid with cell size of 10 nm and is represented by 500 macro electrons and 16 macro ions per cell. A set of initial foil densities ($n_e=80-710~n_c$) and laser intensities ($I=10^{23}-8\times10^{23}~\rm{W/cm^2}$) were used in these simulations.

\section{enhanced $\textnormal\emph{$e^-e^+$}$ pair production}
\vspace{-1 em}

When two counter-propagating laser pulses irradiate on a thin foil from opposite sides, laser hole-boring (HB) stage is initiated. At this stage, the foil electrons and ions are first accelerated in the laser field. Owing to the opposite and equal laser radiation pressures, these accelerated charged particles could hardly be escaped from the foil. The foil then becomes much denser with much higher reflectivity of the lasers, enabling formation of SW on each side of the foil. When HB stage ends, the foil starts thermal expansion stage, in which the sum of electrostatic and thermal pressures may exceed the laser radiation pressure on each side, leading to decompression and expansion of the foil \cite{H. X. Chang2015}. As the electron density of fully ionized foil becomes lower/higher than the relativistically corrected critical density, the foil plasma becomes transparent/opaque to the incident laser \cite{L.Willingale2009}. The results are shown in Figs.~\ref{fig1}(a) and (e). It is seen that the foil plasma is underdense at $n_e=280~n_c$ such that the counter-propagating lasers can penetrate the thin foil [see Fig.~\ref{fig1}(b)] and then form a stable SW. This SW has a period of 0.5$~T_0$. As foil plasma becomes denser, for example, at $n_e=710~n_c$, it can be opaque to the incident laser [see Fig.~\ref{fig1}(f)]. Then the incident and reflected lasers produce instable and relatively weak SWs at both sides of the irradiated foil, as shown in Fig.~\ref{fig1}(e).

Figs.~\ref{fig1}(c) and (g) shows the spatial distributions of hard photons with energies higher than 1.022 MeV. The number of radiated photons increases with the foil plasma density due to proportional increase of foil electron number. These photons have similar spatial patterns with those of foil electrons in the laser focus. At lower foil plasma density, we find that majority of photons are produced with relatively higher energies because the incident lasers can penetrate the thin foil and then enhances the amplitude of EM field. These photons in turn interact with the SW field sufficiently, producing more energetic pairs. The SW field can further bunch the produced pairs in space, leading to the formation of high dense \emph{$e^-e^+$} plasma with a maximum intensity above $10^{20}~\rm{W/cm^2}$, as shown in Fig.~\ref{fig1}(d). Such bunching effect, which disappears in the opaque regime, is mainly caused by the SW formed directly by two incoming laser pulses. When the initial target density is 710$~n_c$, the interaction enters into the opaque regime. The majority of photons are produced at the high-density plasma region, where the incident lasers get reflected [see Fig.~\ref{fig1}(f)]. In the opaque regime, energetic photons experience a relatively weak EM field when they escape from the interaction area, and then \emph{$e^-e^+$} pair production via multi-photon BW process becomes deficient. Fig.~\ref{fig1}(h) shows that the intensity of created positrons at $n_e=710~n_c$ is visibly lower than that at $n_e=280~n_c$. Such intensity profile further displays a chaotic pattern due to the lack of stable SW.

\begin{figure}
  \centering
  \includegraphics[width=9cm]{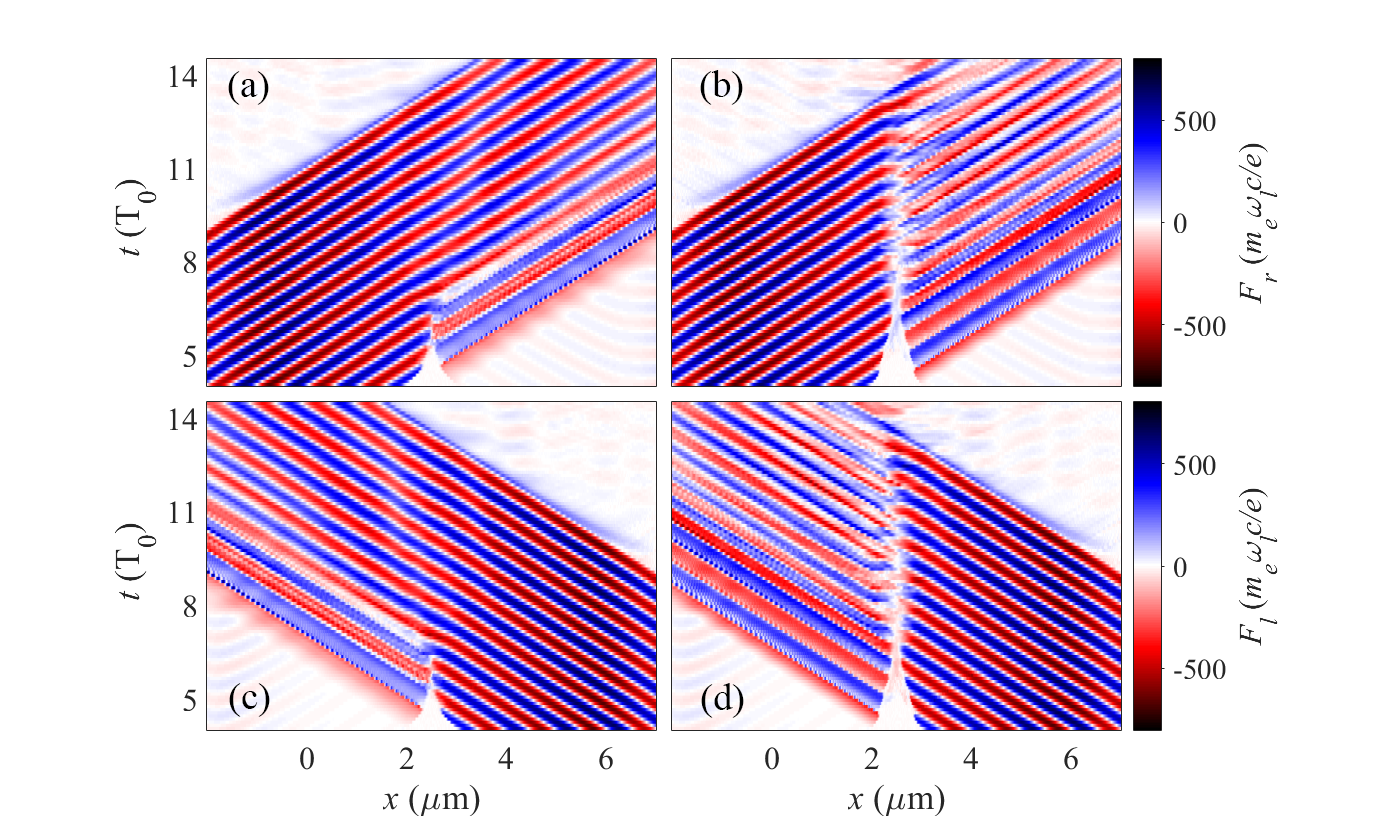}
  \caption{(Color online) Space-time distributions of the right- [(a) and (b)] and left-moving [(c) and (d)] components of the total field for two counter-propagating LP lasers with dimensionless laser amplitude $a_0=540$. The foil densities used in the simulations are $n_e=280~n_c$ [left pads] and $n_e=710~n_c$ [right pads], respectively.}\label{fig2}
\end{figure}

In order to identify the relativistic transparency and opaque regimes, the characteristic behaviors of the right- and left-moving field components are shown in Fig.~\ref{fig2}, which is obtained in simulations with laser amplitude $a_0=540$. These components are defined by \cite{C. Baumann2016}
\begin{equation}\label{1}
  F_{r,~l}=\frac{E_y\pm B_z}{2}.
\end{equation}
One can clearly see that in the case of $n_e=280~n_c$, the onset of transparency begins at $t\thickapprox6~T_0$, which leads to a transient SW [see Figs.~\ref{fig2}(a) and (c)]. The velocity of HB front is estimated as $v_{hb}\thicksim0.3c$ and the HB stage should be terminated at $t\approx5.7~T_0$, which is in accordance with the simulation result. However, in the case of $n_e = 710~n_c$ a visible interruption of laser propagation occurs until $t\thickapprox12T_0$ [see Figs.~\ref{fig2}(b) and (d)], which indicates that the plasma remains opaque for this interaction stage. In the following stage, the injection of laser energy is almost finished and the reminder of the SWs formed directly by two incident lasers can hardly affect the production of \emph{$e^-e^+$} pairs. This is the also the reason that the number of pairs obtained at higher foil density is visibly less than at lower foil density.

\begin{figure}
  \centering
  \includegraphics[width=9cm]{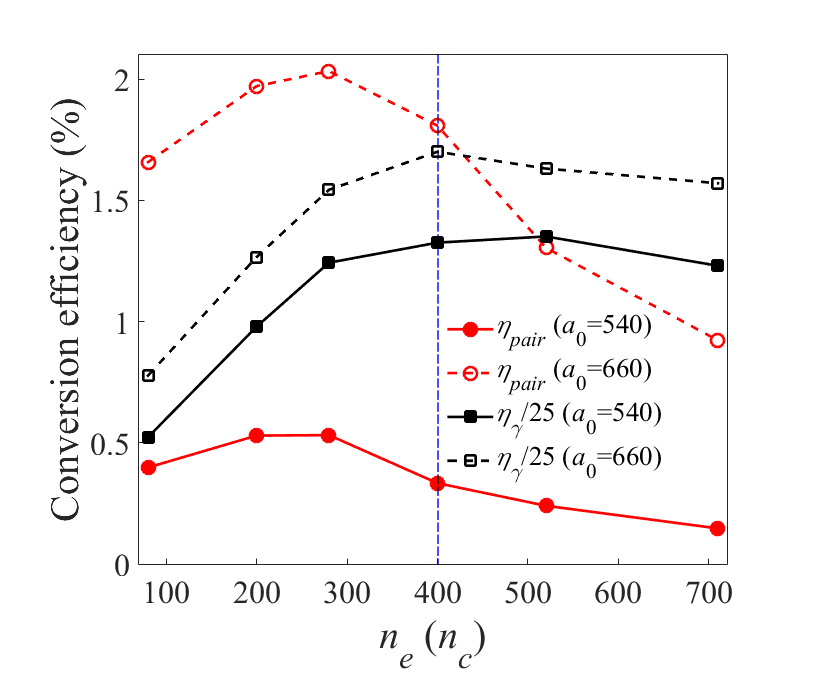}
  \caption{(Color Online) Energy conversion transfer from laser photons to \emph{$e^-e^+$} pairs (cycle) and $\gamma$-photons (square) at $t=13.25~T_0$. Solid lines and dashed lines indicate the simulation results at laser intensities of $4\times10^{23}~\rm{W/cm^2}$ ($a_0=540$) and $6\times10^{23}~\rm{W/cm^2}$ ($a_0=660$), respectively. Vertical line denotes an approximate separatrix between the relativistic transparency and the opaque regimes.}\label{fig3}
\end{figure}

We investigate laser energy conversion to \emph{$e^-e^+$} pairs ($\eta_{pair}$) and $\gamma$-ray photons ($\eta_{\gamma}$) in the relativistic transparency and opaque regimes. Fig.~\ref{fig3} shows such energy transfer as a function of foil plasma density at $t=13.25~T_0$, which corresponds to the moment that the laser fields have vanished. It is seen that the $\eta_{pair}$ increases first and then decreases rapidly with the foil density, whereas the $\eta_{\gamma}$ increases first and then approaches a saturation value. For both laser amplitudes, $a_0=540$ and 660, the $\eta_{pair}$ has a maximum value at foil plasma density of $n_e=280~n_c$. Such density is slightly lower than the separatrix density, approximately given by $n_e=400~n_c$, for which the relativistic transparency occurs at $t\gtrsim9.5T_0$, that is, more than half of the laser pulses have reached the target. The separatrix density observed in the simulations is lower than the theoretical $a_0n_e$, due to target compression and QED plasma effects. The QED plasma effect can deplete the laser pulse energy, and thus reduces the laser amplitude in laser foil interactions. For laser amplitude $a_0=660$, the target compression and laser depletion due to QED effect \cite{W. M.Wang} increase significantly such that the separatrix density does not shift to higher value, compared to that for laser amplitude $a_0=540$.

\begin{figure*}
  \centering
  \includegraphics[scale=0.09]{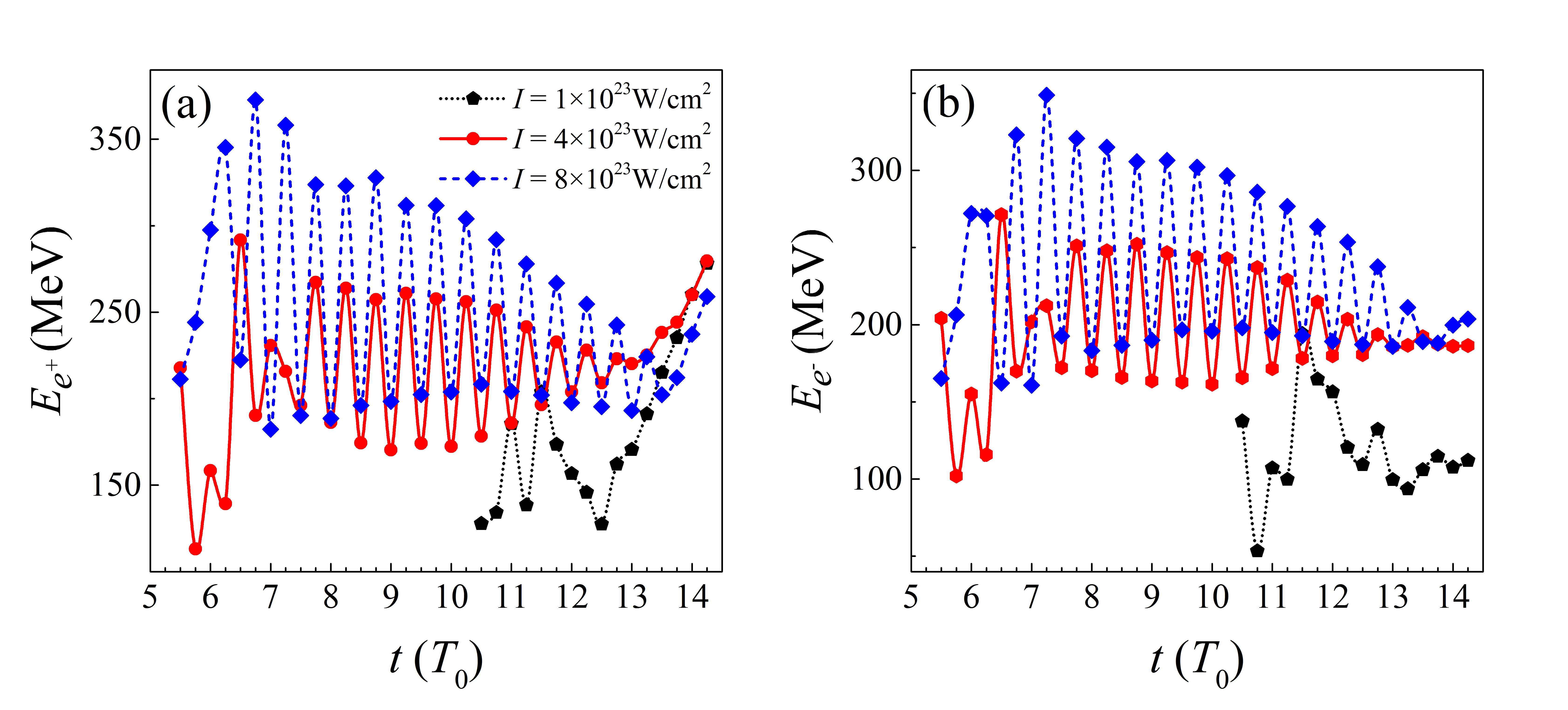}
  \caption{(Color online) Ensemble-averaged energy oscillation for positrons (a) and BW-electrons (b). Two counter-propagating laser pulses at three different intensities $I=1.0\times10^{23}~\rm{W/cm^2}$ (black square), $I=4\times10^{23}~\rm{W/cm^2}$ (red cycle) and $I=8\times10^{23}~\rm{W/cm^2}$ (blue triangle) were used to irradiate a 1$\mu m$ CH foil ($n_e=280~n_c$) from opposite sides.}\label{fig4}
\end{figure*}

As displayed in Fig.~\ref{fig3}, the $\eta_{pair}$ obtained in the relativistic transparency regime is obviously higher than that in the opaque regime. For laser amplitude $a_0=540$, when reducing foil plasma density from 710$~n_c$ to 280$~n_c$ the laser energy conversion $\eta_{pair}$ increases from $0.15\%$ to $0.53\%$, which is enhanced to about four times, so does the number of created pairs $N_{pair}$, from $3.4\times10^{10}$ at $n_e=710~n_c$ to $1.2\times10^{11}$ at $n_e=280~n_c$. Meanwhile, the laser energy conversion $\eta_{\gamma}$ keeps almost the same value, $\thicksim30\%$. The enhancement of \emph{$e^-e^+$} pair production is mainly due to the target transparency. As discussed above, the transparency results in the formation of stable SW from a direct overlap of two counter-propagating laser pulses. Such pulse overlapping enhances the laser field strength, which can accelerate charged particles to higher Lorentz factors and then radiate more high-energy photons per each charged particle. In the following stage, the propagation of these energetic photons through the stable SW field increases the dynamical quantum parameter $\chi_{\gamma}$ \cite{Wen Luo2015,H. X. Chang2015}, which controls \emph{$e^-e^+$} pair production via multi-photon BW process. As a consequence, more QED pairs can be produced in foils undergoing relativistic transparency. We see in Fig.~\ref{fig3} that an optimum foil density region, namely, $200-280~n_c$ is found for realizing maximum conversion efficiency from laser photons to \emph{$e^-e^+$} pairs.
\vspace{-1 em}
\section{modulation dynamics of $\textnormal\emph{$e^-e^+$}$ pair plasma in the relativistic transparency regime}
\vspace{-0.5 em}
\subsection{Energy modulation}
\vspace{-0.5 em}
After transparency has occured, the stable SW can modulate the average pair energy. Fig.~\ref{fig4} shows temporal evolutions of ensemble-averaged energy for the created pairs at laser intensities of $4~\times~10^{23}$ and $8~\times~10^{23}~\rm{W/cm^2}$, respectively. At the initial stage, the HB and thermal expansion effects are dominant over the formation of stable SW. The energy modulation is ambiguous. At $7.5~T_0<t\lesssim12~T_0$, the plasma foil becomes transparent and stable SW fiels are formed consequently [see Figs.~\ref{fig2}(a) and (c)]. During this stage, the average positron energy is modulated periodically [see Fig.~\ref{fig4}]. The modulation period is 0.5$~T_0$, as a result of temporal evolution of the SW. On the contrary, the oscillation of average positron energy becomes insignificant when laser foil interaction occurs in the opaque regime, for example, in the case of $n_e=710n_c$ for $I=4~\times~10^{23}~\rm{W/cm^2}$. This is due to merely the existence of instable SW fields formed by incident and reflected laser pulses. Furthermore, the foil does not undergo transparency at lower intensity of $10^{23}~\rm{W/cm^2}$. Only a few \emph{$e^-e^+$} pairs are produced and the modulation dynamics are invisible as well.

We present detailed modulation processes as follows. When SW field becomes strong at $t=nT_0/2$ with $n$ being the integer, charged particles experience such field and radiate more energetic photons. As a result, average energy per particle decreases and these particles can be readily trapped to the electric nodes due to stronger radiative trapping (RT) \cite{L. L. Jiprl2014} and tight longitudinal confinement, as discussed later. On the contrary, when the overlap of the SWs diminishes at $t=(2n+1)T_0/4$, charged particles radiate less; instead they can absorb some energy from the varying laser fields, which finally enhance their average energy. Such energy oscillation can maintain a few laser periods. After $t\thicksim12~T_0$, it becomes weak and even disappears along with the fading of SW field. It is seen that the final average energy for BW-electrons approaches a median value, approximately 200 MeV. Meanwhile the positron average energy has considerable enhancement on acceleration of sheath potential generated by fast electrons as they leave the target. The existence of sheath potential also gives rise to the average energy difference ($\thicksim$20 MeV) between the positron and BW-electron’s in the oscillation stage, as displayed in Figs.~\ref{fig4} (a) and (b).

The energy oscillation of \emph{$e^-e^+$} pairs can be explained with conservation of canonical momentum of positrons (or BW-electrons) after they are spontaneously accelerated to relativistic velocities. The conservation of canonical momentum reads \cite{J. Meyer-Ter-Vehn192}
\begin{equation}\label{2}
  p_{\perp}+qA_{\perp}/c=constant
\end{equation}

\noindent Here $p_{\perp}$ is the transverse momentum of positron, $q$ is the elementary charge, $A_{\perp}$ is the vector potential of EM field perpendicular to particle momentum, and the $\it{constant}$ can be approximated as the median value mentioned above. During the interaction, the peak amplitudes of the SW field varied periodically, leading to that the trapped pairs have visible oscillation in the average transverse momentum. Meanwhile the longitudinal momentum keeps almost unchanged. The simulations have verified this behavior. A detailed discussion will be given in the following subsection.

At higher laser intensity of $8~\times~10^{23}~\rm{W/cm^2}$, regular modulation comes earlier than at lower laser intensity, as displayed in Fig.~\ref{fig4}. The reason is that the relatively strong laser radiation pressure can shorten the stages of both laser HB and target thermal expansion, hence forming stable SW sooner. Furthermore, the overall charged particle energy decreases due to accumulation of considerable radiation loss, which cannot be compensated by the energy gain from the laser fields.
\vspace{-1.1 em}
\subsection{Phase-space modulation}
\begin{figure}
  \centering
  \includegraphics[width=9cm]{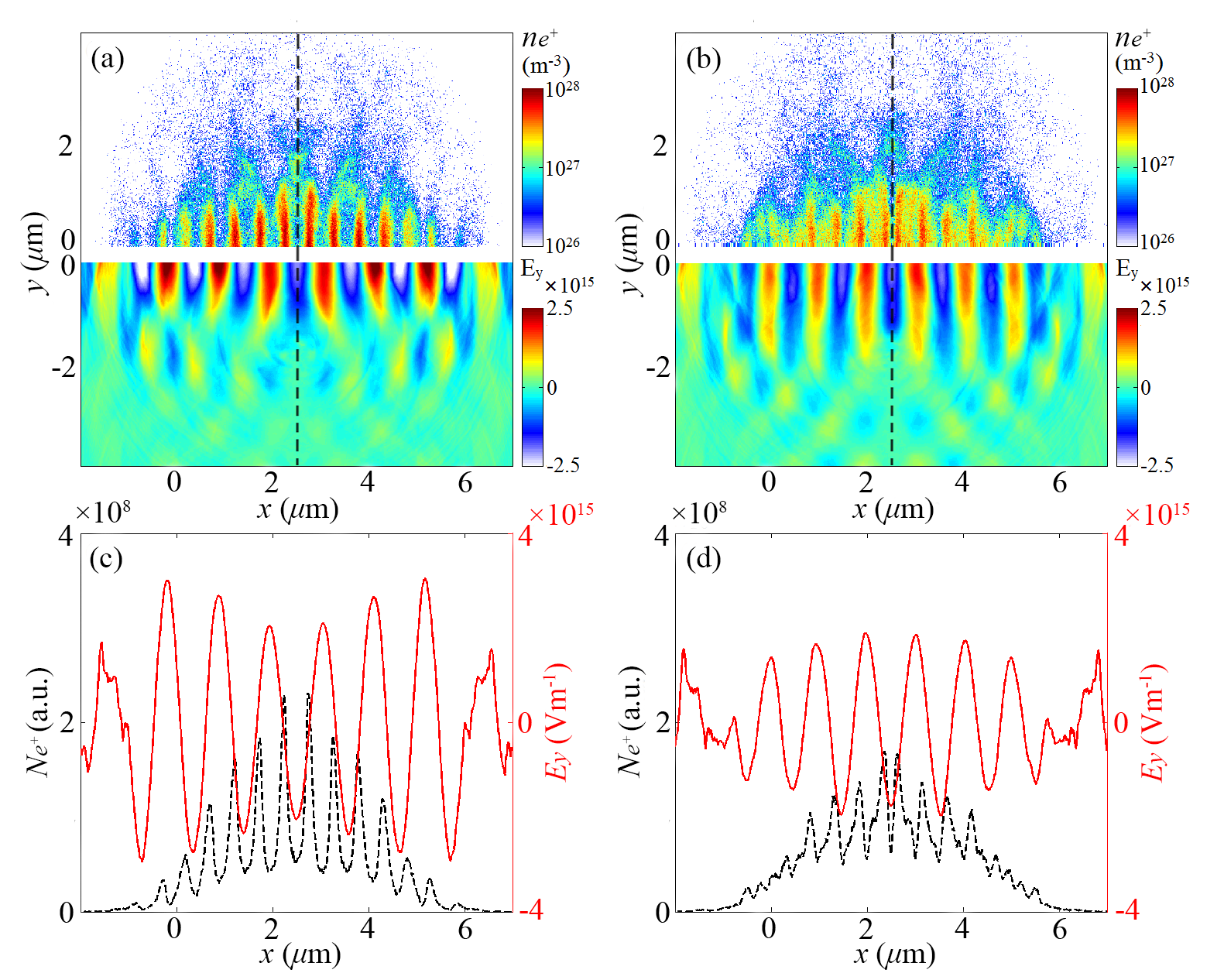}
  \caption{Positron spatial density distributions in $\it{x-y}$ plane (upper parts) as well as longitudinal profiles of electric fields $E_y$ (lower parts) at $t=10~T_0$ (a) and $t=10.25~T_0$ (b), respectively. Due to their mirror patterns along the $\it{x}$-axis, only spatial density profile of positrons in the range of $0<y<4~\mu m$ and longitudinal profile of laser fields $E_y$ in $-4~\mu m<y<0$ are plotted. Longitudinal distributions of positron number (black) and longitudinal profiles of $E_y$ (red) at $t=10~T_0$ (c) and $t=10.25~T_0$ (d). A laser intensity of $4\times10^{23}~\rm{W/cm^2}$ and a foil density of $n_e=280~n_c$ are used in the simulations.}\label{fig5}
\end{figure}

\vspace{-0.5 em}
Besides particle energy oscillation, a visible phase-space modulation can be observed when thin foils undergo relativistic transparency. It is seen in Fig.~\ref{fig5} that the strength of SW field is always zero at nodes of $x=n\lambda_l/2$, and maximum at antinodes of $x=(2n+1)\lambda_l/2$. The field strengths at antinodes alternate between strong and weak overlaps during each half of laser period. The created \emph{$e^-e^+$} pairs are trapped to the nodes of the SW field when the field strength becomes strong at $t=nT_0/2$, which further helps forming stripe distributions along the direction of laser polarization [see an example shown in Fig.~\ref{fig5}(a)]. The \emph{$e^-e^+$} pair bunches have regular space interval of $\lambda_l/2$. After one-fourth of laser period, the bunched pattern evolves into a chaotic lobe when the field strength diminishes [see an example in Fig.~\ref{fig5}(b)].

\begin{figure}
  \centering
  \includegraphics[width=9cm]{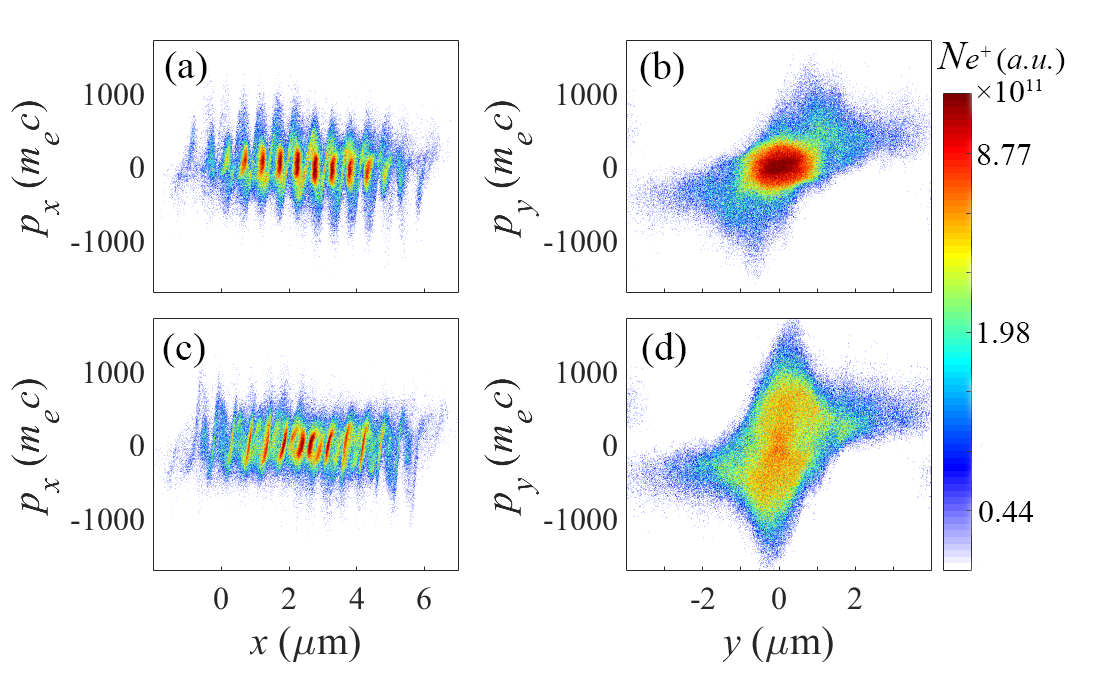}
  \caption{(Color online) Positron phase-space distributions for $x-p_x$ at time $t=10~T_0$ (a) and $t=10.25~T_0$ (c), and for $y-p_y$ at $t=10~T_0$ (b) and $t=10.25~T_0$ (d).}\label{fig6}
\end{figure}

\begin{figure*}
  \centering
  \includegraphics[scale=0.6]{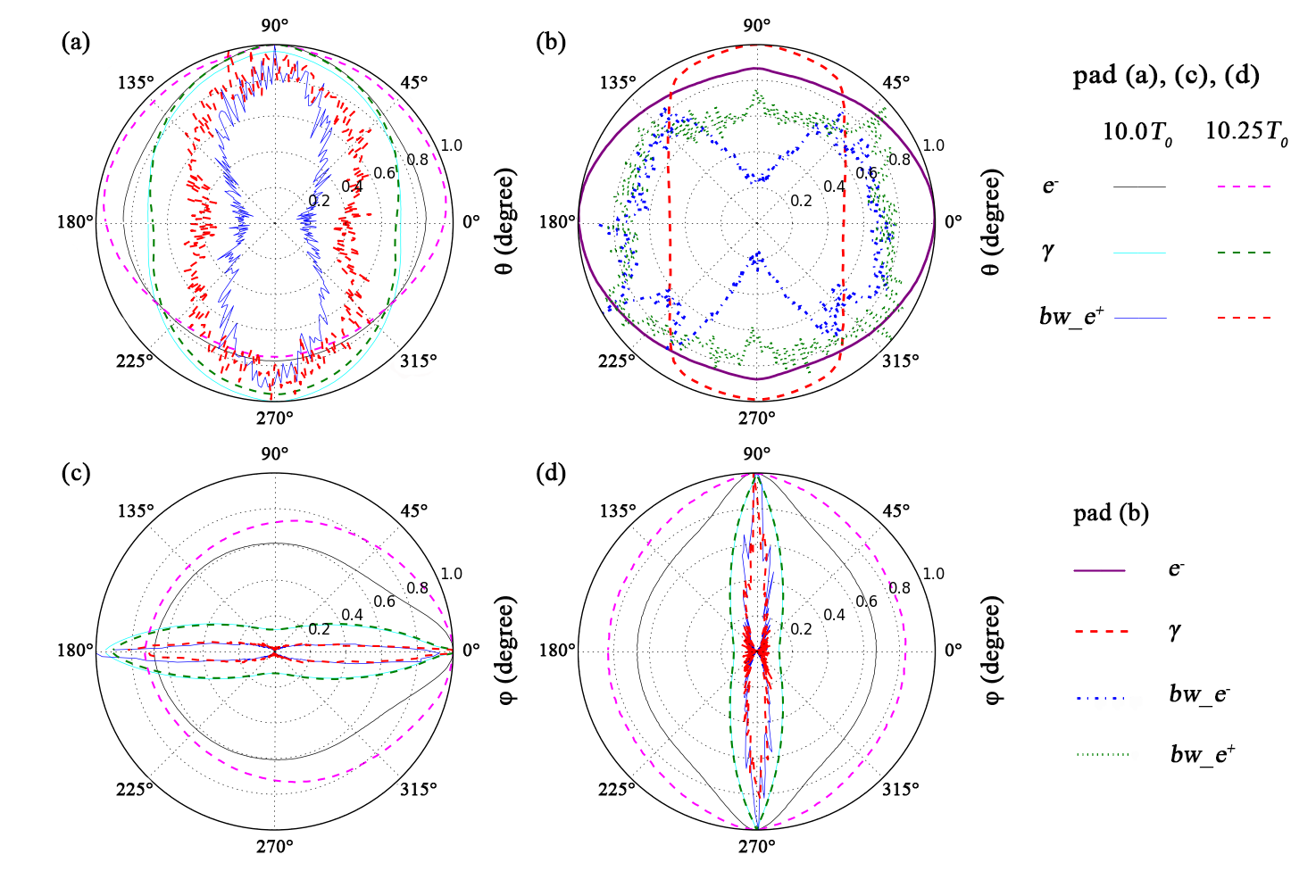}
  \caption{(Color online) Polar distributions of foil electron, $\gamma$-photon and positron intensities at time $t=10.0~T_0$ and 10.25$~T_0$ (a) and polar distributions of foil electron, $\gamma$-photon, BW-electron and positron intensities at $t=14.5~T_0$ (b), for $\it{p}$-polarized laser. Azimuthal distributions of foil electron, $\gamma$-photon and positron intensities at $t=10.0~T_0$ and 10.25$~T_0$, for $\it{p-}$ (c) and $\it{s-}$ (d) polarized lasers. In Pads (a), (c) and (d), the black, cyan and blue solid lines represent foil electrons, $\gamma$-photons and positrons, respectively at $t=10.0~T_0$; the magenta, green and red dashed lines indicate those at $t=10.25T_0$, respectively; the polar distribution of BW-electron intensity displays similar pattern as the positron’s, thus it is not shown here. Laser intensity of $4~\times~10^{23}~\rm{W/cm^2}$ and foil density of $n_e=280~n_c$ were used in the simulations.}\label{fig7}
\end{figure*}

The positron density peaked at electric nodes can be clearly seen in Fig.~\ref{fig5}(c), which plots the longitudinal distribution of positron number for a fixed moment in time $t=10~T_0$. At this moment, charged particles experience an electric field almost doubles that they experience at $t=10.25~T_0$ [see Fig.~\ref{fig5}]. Consequently these particles quiver violently in the intense field and loose substantial amount of energy by radiating high-energy photons. They in turn experience strong radiation-reaction (RR) force, leading to visible RT. A direct evidence for RT dynamics is the decrease of trapped particle transverse momentum \cite{L. L. Jiprl2014}, which is visibly shown in Fig.~\ref{fig6}. Instead of being focused only in the electric nodes, more positrons tend to appear in the subspace between two electric nodes at $t=10.25~T_0$ [see Fig.~\ref{fig5}(d)].

Figs.~\ref{fig6}(a) and (b) show positron phase-space distributions at time $t=10~T_0$. The $x-p_x$ distribution has transverse stripe structure periodically, which agrees with its spatial pattern shown in Fig.~\ref{fig5}(a). Instead of being expelled transversely by the ponderomotive force, positrons are mainly trapped in the high-field region of SW. Consequently the $y-p_y$ distribution in Fig.~\ref{fig6}(b) displays in a bright elliptical pattern in the center. However, the particle dynamics is different at time $t=10.25~T_0$, which corresponds to that the field strength of SW at their antinodes. At this moment both the confinement of charged particles and RT effect loose. These positrons with $p_x>0$ tend to move forward, and those with $p_x<0$ move backward. As a result, the transverse stripe structure becomes oblique with respect to the longitudinal direction [see Fig.~\ref{fig6}(c)]. The weakened SW field further disperses the positrons in phase-space of $y-p_y$, since the RR effect becomes insignificant such that it is difficult to compensate their transverse dispersion.
\vspace{-1.25 em}
\subsection{Angular distribution}
\vspace{-0.5 em}
We continue to investigate angular distributions of foil electrons, $\gamma$-photons and \emph{$e^-e^+$} pairs. Since it is necessary to consider additional plasma effects, such as thermal expansion along $\it{z}$-direction, more realistic three-dimensional (3D) simulations were performed. We define $\theta=0^{\circ}$ along the direction of the laser propagation, and $\phi=0^{\circ}~(90^{\circ})$ along the direction of $\it{p (s)}$-polarization. Two-side irradiation scheme enables irradiating and compressing a thin foil more symmetrically, thus foil electron, $\gamma$-photon and positron have symmetric $\theta$-intensity patterns with respect to the azimuthal plane [see Fig.~\ref{fig7}(a)]. Since the laser intensity is much higher than the relativistic intensity of the order of $10^{18}~\rm{W/cm^2}$, the effect of the magnetic and electric fields on the motion of charged particles should become comparable. Relativistic \emph{$e^-e^+$} pairs produced in such field are predicted to quiver nonlinearly, moving in figure-of-eight trajectory, rather than in straight lines.

The modulation of polar distribution of positron intensity is seen from Fig.~\ref{fig7}(a). In the presence of SW field, the polar patterns of positrons at time $t=10.0~T_0$ and 10.25$~T_0$ are “pinched” along the direction of laser polarization due to the confinement effect of the SW field. The polar pattern at time $t=10.0~T_0$ becomes more slim relative to that at $t=10.25~T_0$. This is due to a tighter confinement effect longitudinally. In order to confirm such effect, we display in Fig.~\ref{fig7}(b) that the 'pinched' effect disappears when the SW field has already faded. It is also shown in Fig.~\ref{fig7}(b) that different from BW-electron, the positron intensity have an isotropic distribution, due to the influence of sheath acceleration. For $\it{p}$-polarized lasers, foil electrons experience transverse electric fields. They and hard photons they emit have tendency to move along the direction of laser polarization, $\it{i.e}$., $\theta=90^{\circ}$.

Figs.~\ref{fig7}(c) and (d) show azimuthal distributions of foil electron, $\gamma$-photon and positron intensities for $\it{p-}$ and $\it{s}$-polarized lasers, respectively. In both laser polarization cases, foil electrons have angular patterns with maximal intensity in their respective laser polarization directions. However, such pattern is a little ‘fat’ as a result of target thermal expansion after it symmetrically compressed by laser radiation pressures. Azimuthal distribution of $\gamma$-photon intensity is peaked on-axis of the laser polarization, suggesting high-order harmonic generation based on Compton scattering in highly nonlinear regime. The positron intensity displays similar patterns as the $\gamma$-photon. Also shown in Figs.~\ref{fig7}(c) and (d) is temporal evolutions of SW field can hardly affect positron azimuthal pattern, which is different from modulation dynamics on positron polar structure.
\vspace{-1.25 em}
\section{summary and conclusion}
\vspace{-0.5 em}
Thin foil undergoing relativistic transparency can enhance the \emph{$e^-e^+$} pair production. The produced pair plasma has a density as high as $10^{22}~\rm{cm^{-3}}$. This implies the beam proper density of $n_{prop}=n_e/\gamma_{AV}\simeq2.5\times10^{19}~\rm{cm^{-3}}$ while considering the bulk Lorentz factor of $\gamma_{AV}\simeq400$. The beam of the produced pair plasma has transverse size of the order of $D_B\simeq2~\mu m$. The relativistic corrected collision-less skin depth of the beam is thus $l_{skin}=c/\omega_{prop}\simeq1.1~\mu m$, with $\omega_{prop}$ being the relativistic plasma frequency. This value is smaller than the beam transverse size, indicating that collective (that is, plasma-like) behaviors are likely to occur in the beam. Since the created plasmas are dense enough to permit collective and kinetic behaviors to play roles, they are further similar to the condition of astrophysical events such as jets of long $\gamma$-ray bursts \cite{J. Wardle1998,I. F. Mirabel1999}.

In summary, the generation of overdense pair plasmas and following modulation dynamics have been investigated by simulations of two QED-strong lasers irradiating a thin foil from opposite sides. In the relativistic transparency regime the laser energy conversion to \emph{$e^-e^+$} pairs can increase four times compared to that in the opaque regime, thus demonstrating enhanced \emph{$e^-e^+$} pair production. At laser intensity of $4~\times~10^{23}~\rm{W/cm^2}$, the produced \emph{$e^-e^+$} pair plasmas have a high energy density exceeding $10^{20}~\rm{W/cm^2}$. Although the beam transverse size is only a few micros, the created pair plasmas are dense enough to allow collective and kinetic behaviors to play roles. The modulation dynamics of \emph{$e^-e^+$} pairs is further demonstrated after transparency has occurred, which shows that the positron average energy, phase-space and angular distributions can be modulated periodically by the stable SW field formed directly by the counter-propagating laser pulses.
\vspace{-1.25 em}
\section{acknowledgements}
\vspace{-0.5 em}
This work was supported by the National Natural Science Foundation of China (Grant Nos. 11405083, 11347028 and 11675075), and the Graduate Student Innovation Project Foundation of Hunan Province (Grant No. CX2016B450). W.L. appreciates the support from China Scholarship Council and Young Talent Project of the University of South China. M.C. appreciates the support from National 1000 Youth Talent Project of China.

\end{document}